\documentclass{jpp}

\usepackage{graphicx}
\usepackage{natbib}
\usepackage{color}

\ifCUPmtlplainloaded \else
  \checkfont{eurm10}
  \iffontfound
    \IfFileExists{upmath.sty}
      {\typeout{^^JFound AMS Euler Roman fonts on the system,
                   using the 'upmath' package.^^J}%
       \usepackage{upmath}}
      {\typeout{^^JFound AMS Euler Roman fonts on the system, but you
                   dont seem to have the}%
       \typeout{'upmath' package installed. JPP.cls can take advantage
                 of these fonts, if you use 'upmath' package.^^J}%
      }
  \else
  \fi
\fi

\ifCUPmtlplainloaded \else
  \checkfont{msam10}
  \iffontfound
    \IfFileExists{amssymb.sty}
      {\typeout{^^JFound AMS Symbol fonts on the system, using the
                'amssymb' package.^^J}%
       \usepackage{amssymb}%

      }{}
  \fi
\fi

\ifCUPmtlplainloaded \else
  \IfFileExists{amsbsy.sty}
    {\typeout{^^JFound the 'amsbsy' package on the system, using it.^^J}%
     \usepackage{amsbsy}}
    {}
\fi

\newsavebox{\astrutbox}
\sbox{\astrutbox}{\rule[-5pt]{0pt}{20pt}}

\title[PCX]{Prospects for observing the magnetorotational instability in the Plasma Couette Experiment}

\author[K. Flanagan et al.]%
{K.\ns F\ls L\ls A\ls N\ls A\ls G\ls A\ls N$^1$%
  \thanks{Email address for correspondence: ksflanagan@wisc.edu},\ns
M.\ns C\ls L\ls A\ls R\ls K$^{1}$,\ns C.\ns C\ls O\ls L\ls L\ls I\ls N\ls S$^{1,2}$,\ns C.\ns M.\ns C\ls O\ls O\ls P\ls E\ls R$^{1}$,\ns\break
I.\ns V.\ns K\ls H\ls A\ls L\ls Z\ls O\ls V$^{1,3}$,\ns J.\ns W\ls A\ls L\ls L\ls A\ls C\ls E$^{1}$\ns \and C.\ns B.\ns F\ls O\ls R\ls E\ls S\ls T$^{1}$}

\affiliation{$^1$Department of Physics, University of Wisconsin, Madison, WI 53706, USA\\[\affilskip]
		$^2$ University of California Irvine, Irvine, CA 92697, USA\\[\affilskip]
		$^3$National Research Centre ``Kurchatov Institute", Moscow, 123182, Russia\\[\affilskip]}

\pubyear{2010}
\volume{650}
\pagerange{119--126}
\begin{document}

\maketitle

\begin{abstract}
Many astrophysical disks, such as protoplanetary disks, are in a regime where non-ideal, plasma-specific magnetohydrodynamic (MHD) effects can significantly influence the behavior of the magnetorotational instability (MRI). The possibility of studying these effects in the Plasma Couette Experiment (PCX) is discussed. An incompressible, dissipative global stability analysis is developed to include plasma-specific two-fluid effects and neutral collisions, which are inherently absent in analyses of Taylor-Couette flows (TCFs) in liquid metal experiments. It is shown that with boundary driven flows, a ion-neutral collision drag body force significantly affects the azimuthal velocity profile, thus limiting the flows to regime where the MRI is not present. Electrically driven flow (EDF) is proposed as an alternative body force flow drive in which the MRI can destabilize at more easily achievable plasma parameters.  Scenarios for reaching MRI relevant parameter space and necessary hardware upgrades are described.
\end{abstract}


\section{Introduction}
The magnetorotational instability (MRI), first derived by \citet{VelikhovMRI} and \citet{ChandrasekharMRI}, has been studied extensively as a possible mechanism of enhanced angular momentum transport in accretion disks \citep{BalbusHawley91,BalbusReview}. In an idealized form with no dissipative mechanisms, this instability occurs when conducting fluid in a spinning disk is threaded by a weak magnetic field. In order to excite the instability the fluid must have an angular velocity profile which is decreasing with radius, $\partial\Omega^{2}/\partial\ln{r}<0$, and the magnetic field must be sufficiently weak as to not stabilize perturbations, $({\bf k}\cdot{\bf V}_{A})^{2}<-\partial\Omega^{2}/\partial\ln{r}$. Dissipation enters through both the fluid viscosity, $\nu$, and the magnetic diffusivity, $\eta$. These dissipative mechanisms are commonly parameterized by the fluid and magnetic Reynolds numbers: $Re=VL/\nu$ and $Rm=VL/\eta$, respectively. When $Re$ is small, fluid viscosity acts to dampen any instabilities, thus there is a minimum $Re$ threshold for observing the MRI. Likewise, when the magnetic field advection is small compared to diffusion, i.e. $Rm$ is small, magnetic field lines are not affected by flows and there is no feedback mechanism for the MRI, thus a minimum $Rm$ value must be met as well. 

The ratio between fluid viscosity and magnetic diffusivity is also key to understanding the MRI. This ratio is described by the magnetic Prandtl number, $Pm=Rm/Re=\mu_{0}\nu/\eta$ and can vary by many orders of magnitude in different astrophysical systems: $Pm\lesssim10^{-5}$ in disks around young stellar objects while $Pm\gtrsim10^5$ around active galactic nuclei. Shearing box simulations have shown that varying $Pm = 0.06-4$ can cause the magnitude of the MRI driven turbulent radial momentum transport to vary by two orders of magnitude \citep{Longaretti_2010}, suggesting that the MRI may not be efficient at small Prandtl numbers. Unfortunately for liquid metal experiments, $Pm$ is fixed at $\sim10^{-5}$, thus making them susceptible to Ekman circulation driven fluid turbulence (caused by large $Re$) and very small saturated amplitudes (caused by small $Rm$) \citep{Gissinger2012}. However, in plasmas $Pm\propto n^{-1}T_{e}^{3/2}T_{i}^{-5/2}$ and can be adjusted by varying the density and temperature. 

\begin{figure}
	\centerline{\includegraphics[scale=1]{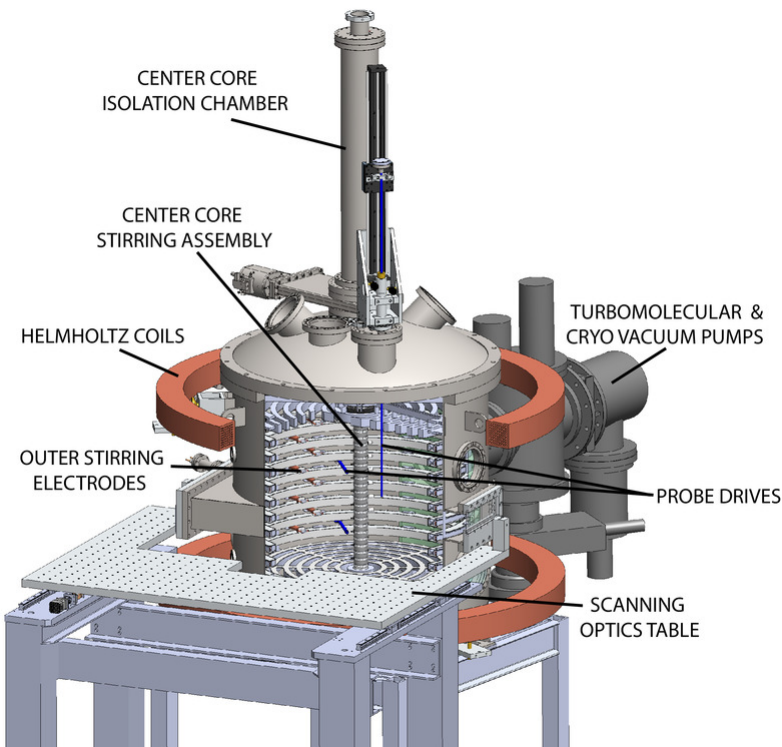}\includegraphics[scale=1]{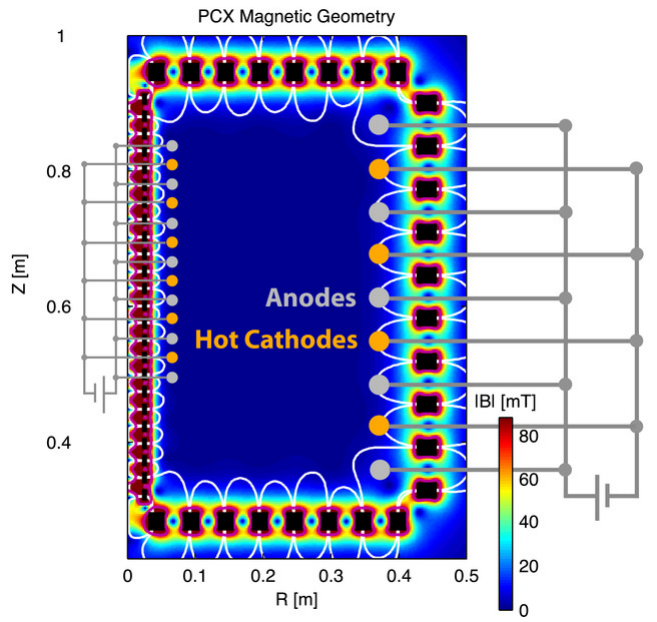}}
	\caption{Left: A cutaway view of PCX showing the major subsystems. Right: The magnetic field geometry of PCX showing electrodes at both the inner and outer boundaries.}
\label{fig:Figure1}
\end{figure}

Motivated by this control of $Pm$ in plasmas, the Plasma Couette Experiment (PCX) explores the magnetohydrodynamic stability of hot, fast-flowing plasmas in a Taylor-Couette flow (TCF) geometry. In PCX, steady-state plasmas are created and confined in a magnetic field-free volume and spun via electrostatic stirring at inner and outer boundaries of a cylindrical plasma volume \citep{CamiPRL,CamiPoP}. Differential flows with peak velocities up to 12 km s$^{-1}$ in helium, densities of 10$^{11}$ cm$^{-3}$, and $T_{e}\sim8$ eV are routinely created, corresponding to $Rm\sim65$. TCF involves boundary driven flows, where momentum injected at inner and outer boundaries (via rotating walls in liquid metal experiments or electromagnetic drive in PCX) is viscously coupled to the bulk fluid. This is distinctly different from body forces like gravity or electromagnetic forces. To observe the MRI, TCF flow profiles must be adjusted to mimic Keplerian-like (body force) flows of astrophysical accretion disks, where angular velocity is decreasing with radius while still maintaining Rayleigh hydrodynamic stability with angular momentum increasing with radius (the Rayleigh criterion). In PCX, the range of achievable densities and temperatures set $Pm\approx10^{-1}-10^{2}$.

PCX operates at densities of $10^{10}-10^{11}$ cm$^{-3}$, which correspond to an ion inertial length on the order of the vessel size or larger. Under these conditions the Hall effect, where ions become decoupled from the magnetic field, has a strong influence on the MRI. One of the most dramatic consequences of Hall-MRI is the requirement that the magnetic field threading the flow be antiparallel to the axis of rotation \citep{wardle99,balbus01_aj}. Additionally, PCX works at rather low ionization fractions, $f_{\%}\lesssim1\%$ in helium. The high neutral density acts as a retarding body force on the bulk ion flow via charge-exchange collisions and can have dramatic effects on the boundary driven profiles, thus the stability of PCX flows. Recent work by \citet{Kunz2013,Lesur2014} suggests that two-fluid and neutral effects are necessary for qualitatively correct descriptions of the MRI in protoplanetary discs. Because of the importance of two-fluid and neutral collision effects in real astrophysical systems and the adjustable $Pm$, plasma experiments, such as PCX, are ideal for laboratory study of the MRI.  

This article focuses on the effects of dissipation, the Hall term, and the neutral drag body force on the onset of the MRI in PCX. First, a description of PCX and its current operation is provided. Extensive work has been spent understanding the effect of the neutral drag body force on boundary driven flow profiles. In order to mitigate the effects of neutrals on PCX flows, a confinement and input power upgrade, PCX-U, is described. Using the estimated parameters of PCX-U a global stability analysis of the MRI including the neutral drag force and Hall effect is presented. Results show that even in the improved PCX-U, the neutral drag body force can drive hydrodynamic instability of flows, thus the final portion of this article focuses on the equilibrium and stability of a body driven flow scheme which we simply call electrically driven flow (EDF).  
 
\section{Description of Experiment}\label{sec:descrip}

\begin{figure}
  \centerline{\includegraphics[scale=1]{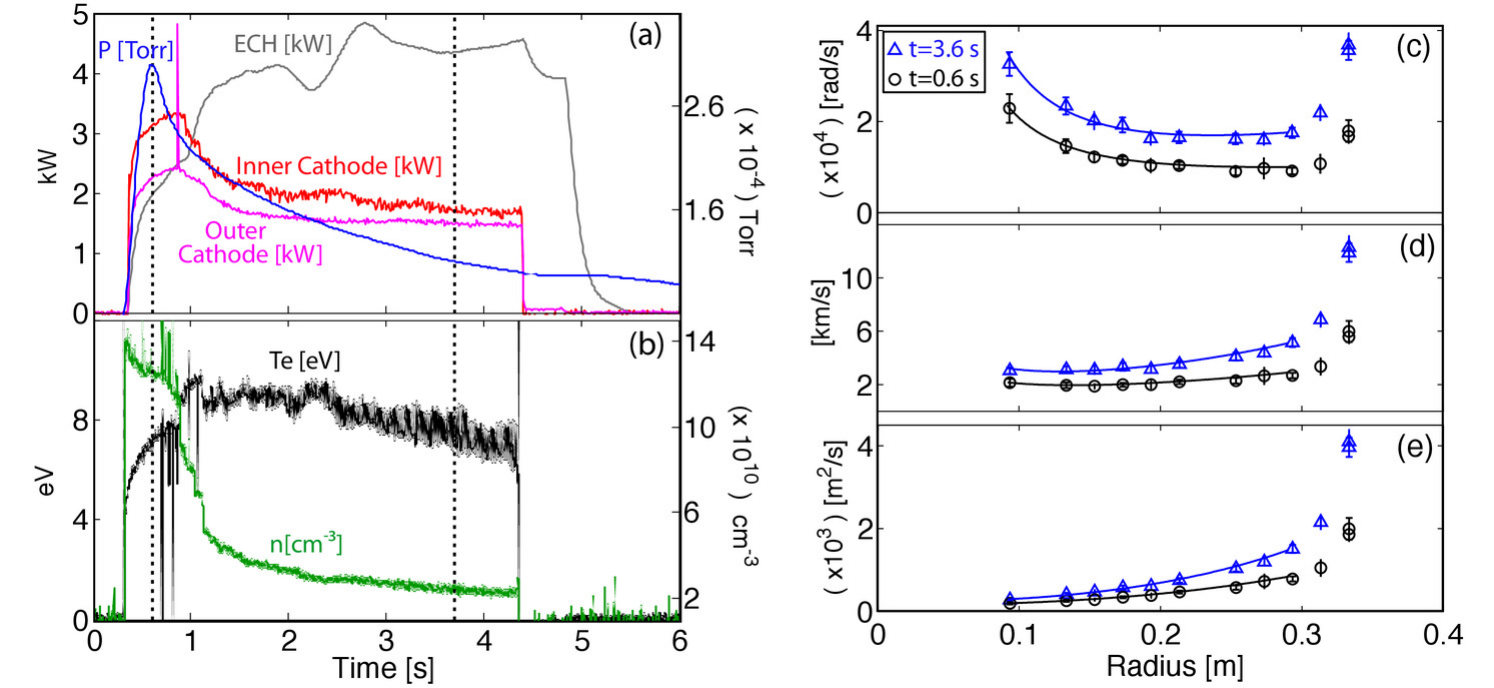}}
  \caption{An example discharge from PCX. In (a) plasma is created with a neutral gas puff and ECH heating. The plasma is spun with four outer cathodes (each 350 V, $\sim1.1$ A) and five inner cathodes (each 575 V, $\sim0.75$ A). In (b) electron temperature and density as measured by the triple Langmuir probe. (c) The angular velocity profile is shown at two different times in the shot (shown as dashed vertical lines in (a) and (b)) along with the azimuthal velocity (d) and angular momentum (e) profiles. The error bars are standard deviation of fluctuation in time.}
\label{fig:Figure2}
\end{figure}

PCX consists of a roughly 1 m tall by 1 m diameter cylindrical vacuum chamber. The inner walls of the chamber are lined with toroidal rings of permanent magnets. This magnet geometry is shown in Fig. \ref{fig:Figure1}. The rings alternate polarity creating a strong multi-cusp magnetic field isolated to the edge of the vessel. This field confines the plasma losses to a small cusp on the face of the magnets, while leaving the bulk volume unmagnetized. If desired, an external Helmholtz coil can produce nearly uniform axial magnetic fields in the chamber up to $B_{0}\sim50$ G. 

A plasma discharge is created by biasing emissive thoriated tungsten cathodes to cold molybdenum anodes at up to 600 V. The plasma is then heated with up to 6 kW of electron cyclotron heating (ECH). Table \ref{tab:params} shows typical PCX plasma parameters along with the relevant dimensionless quantities for this paper, while Fig. \ref{fig:Figure2} shows the time evolution of plasma parameters in a typical PCX discharge.

The hot cathodes also act to stir the plasma by drawing a current across the multi-cusp magnetic field at the edge. This ${\bf J}\times{\bf B}$ torque viscously couples inward to the unmagnetized bulk plasma. Due to the axisymmetric configuration of the magnets, particle drifts in the magnetized edge act to symmetrize the system, such that the toroidal location of the cathodes and anodes does not affect the flow drive \citep{Katz2012RSI}. Flow can be driven at the inner boundary by a center core assembly consisting of 22 stacked magnet rings and up to 8 smaller thoriated tungsten cathodes biased with respect to cold anodes. This assembly can be entirely removed via a gate valve for maintenance of the cathodes. 

\begin{table}
  \begin{center}
\def~{\hphantom{0}}
 \begin{tabular}{l || c | c }
{}&PCX&PCX-U\\ \hline
$n_{e}$ (cm$^{-3}$)&$10^{10}-10^{11}$&$10^{11}-10^{12}$\\
$T_{e}$ (eV)&$5-10$&$10-20$\\
$T_{i}$ (eV)&$0.1-0.5$&$0.5-1$\\
$f_{\%}$&$0.1-10\%$&$1-99\%$\\
$M\equiv V_{1}/C_{s}$&$0.5-0.7$&$0.3-0.5$\\
$M_{A}\equiv V_{1}/V_{A}$&$0-4$&$0-12$\\
$Rm\equiv (V_{1} R_{1})/\eta$&$0-500$&$0-1600$\\
$Pm\equiv Rm/Re$&$0.2-200$&$0.5-350$\\
$Ha\equiv \sqrt{ReRm}/M_{A}$&$40-80$&$30-50$\\
$\mu\equiv R_{1}/L_{\nu}$&$0.1-10$&$0.02-2$\\
$\delta_{i}\equiv c/(\omega_{pi}R_{1})$&$5-50$&$2-15$
\label{tab:params}
\end{tabular}
  \caption{Plasma parameter and dimensionless variable ranges for both PCX and PCX-U. For the Hartmann number ($Ha$) and Alfv\'{e}n Mach number ($M_{A}$), a magnetic field of $B_{0}=-2$ Gauss is used. The Mach numbers used are based on velocity at the inner radius. The unit of length used in this paper is the inner stirring radius $R_{1}=0.1$ m.}
  \label{tab:params}
  \end{center}
\end{table}

Plasmas in PCX are routinely diagnosed using a swept or triple tip Langmuir probe for electron temperature and density, a cold cathode gauge for neutral pressure, and Mach probes for flow velocity. In addition to these routine diagnostics, PCX has implemented Optical Emission Spectroscopy (OES) and a Fabry-Perot interferometer to measure electron and ion temperature, respectively. Both of these optical diagnostics are non-invasive, operating by simply sampling a single chord of emitted plasma light. The OES system takes a broadband spectrum of the plasma and uses line ratios to determine the electron temperature. OES temperature measurements in PCX agree within 15\% of the routine Langmuir probe. The Fabry-Perot interferometer finely samples a small ($<1$ nm) range of the emission spectrum centered at 468.6 nm in helium and 488 nm in argon. Ion emission lines at these wavelengths reflect the entire ion distribution function, such that the ion temperature and velocity can be deduced. In initial measurements of non-flowing argon plasmas, the ion temperature was determined to be $T_{i}=0.2-1$ eV. 

\subsection{Taylor-Couette Flow (TCF) Profiles with Ion-Neutral Drag}\label{sec:TCF}

In weakly-ionized plasmas charge exchange collisions between neutrals and ions impose a body drag force that affects the equilibrium (steady-state) velocity profiles. This neural drag effect has been directly observed in PCX flows \citep{CamiPRL}. An analytical expression for neutral drag modified profiles can be found by treating the neutral drag as a momentum sink in the toroidal momentum balance equation. These profiles are a departure from ideal TCF yet still have a simple Bessel function form:

\begin{equation}\label{eq:TCF}
V_{\phi}(r)=A\mathcal{I}_{1}(r/L_{\nu})+B\mathcal{K}_{1}(r/L_{\nu})
\end{equation}
\\
where the constants $A$ and $B$ depend on the outer and inner boundary locations and the velocities applied, $L_{\nu}^{2}\equiv\tau_{i0}\nu$ is the momentum diffusion length, $\tau_{i0}=(n_{0}<\sigma_{cx}v>)^{-1}$ is the ion-neutral collision time, $\sigma_{cx}$ is the ion-neutral charge exchange cross section, and $\nu$ is the kinematic viscosity due to ion-ion collisions. The momentum diffusion length represents the combined effects of viscous momentum diffusion and the neutral drag momentum sink. When $\mu\equiv R_{1}/L_{\nu}\gg1$, the neutral drag dominates the viscous diffusion, rotation is confined to inner and outer edges of the plasma. These heavily affected profiles have a much lower average velocity across the profile and increased shear near the boundaries that can drive hydrodynamic instability. In the opposite limit $\mu\ll 1$, the velocity profile becomes the ideal Taylor-Couette profile: $V_\phi(r)=ar+b/r$. 

\begin{figure}
  \centerline{\includegraphics[scale=1]{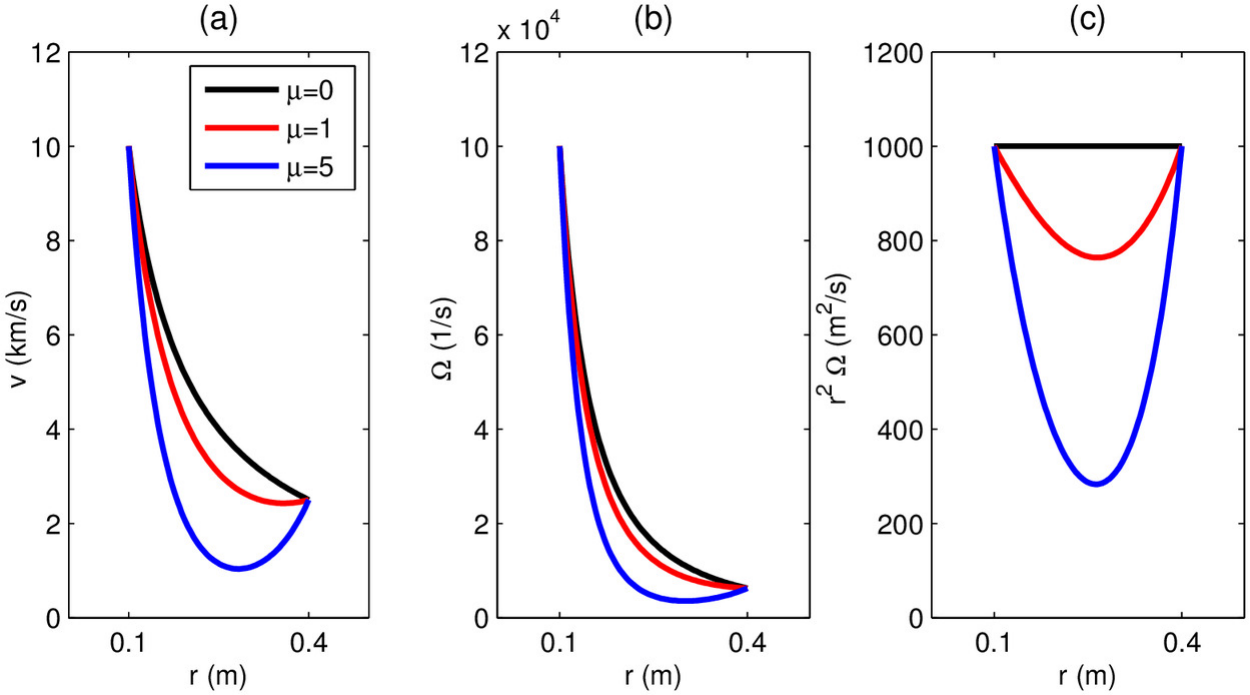}}
  \caption{Synthetic profiles of the (a) toroidal velocity, (b) angular frequency, and (c) angular momentum of modified Taylor-Couette flow with different $\mu\equiv R_{1}/L_{\nu}$ values. The effect of the neutral drag is increased at greater values of $\mu$.}
\label{fig:Figure3}
\end{figure}

Finite neutral drag has a large qualitative effect on the velocity profile of boundary driven flows and greatly affects the stability of this modified TCF. Figure \ref{fig:Figure3} highlights the effect of ion-neutral collisions on TCF profiles of toroidal velocity, angular frequency and angular momentum. The boundary velocities for these profiles are chosen to be $V_{1}=10$ km s$^{-1}$ at $R_{1}=0.1$ m and $V_{1}=2.5$ km s$^{-1}$ at $R_{1}=0.4$ m. With no neutrals this gives a toroidal velocity profile $V_\phi(r)\propto 1/r$, which is marginally stable to the Rayleigh criterion and meets the ideal MRI condition. For heavily affected profiles ($\mu\gg1$), the Rayleigh criterion is not met across the whole profile and hydrodynamic instability is expected. Such a choice of boundary velocities also allows comparison of TCF with electrically driven flow, where $V_\phi(r)\propto 1/r$ is driven via a body force (Sec. \ref{sec:EDF}). 

The profiles shown in Fig. \ref{fig:Figure3} have not been created on PCX due to the current limitations of the centerstack stirring assembly, but they are expected to be achieved in the upgraded experiment. TCF profiles of up to 12.3 km s$^{-1}$ on the outer boundary and 3.4 km $^{-1}$ on the inner have been driven in helium PCX plasmas \citep{CamiPoP}. These flows, shown in Fig. \ref{fig:Figure2}, have areas of decreasing angular frequency as required by the ideal-MRI criteria. They also have increasing angular momentum, thus meet the Rayleigh criterion for hydrodynamic stability. At PCX plasma parameters, these flows correspond to a Reynolds number of $Re\approx26$ (using the outer boundary velocity) and a magnetic Prandtl number of $Pm\approx2.5$. However, these flows have thus far been obtained without applied axial magnetic field so no instabilities have yet been observed.

\subsection{Proposed System Upgrades}\label{sec:PCX_U}

An upgrade to the confinement and heating systems for PCX is underway. The goals are to improve confinement by increasing the volume to loss area ratio, $\mathbb{V}/A_{\ell}$, to input more power, and to improve the temperature tolerances of PCX in order to accommodate longer and higher power discharges. These upgrades, referred to as PCX-U from here on, constitute a major change for PCX that collectively aims to push the experiment into MRI relevant parameter regimes. Expected parameters are presented in Table~\ref{tab:params}. 

The largest portion of the PCX upgrade consists of an entirely new magnet assembly. Approximately 2000 samarium cobalt (SmCo) magnets, like those used in the Madison Plasma Dynamo Experiment (MPDX) \citep{MPDXPOP}, will replace the current array of ceramic magnets. These relatively inexpensive magnets have a much higher field strength and temperature tolerance. For the sizes and grade proposed, the field will be approximately 4 kG at the face of the magnets (the ceramic magnets are 1 kG) and the maximum operating temperature will be nearly 300$^{\circ}$C. 

The new magnets themselves will be a significant improvement over the current system. Currently, the length of PCX discharges is mostly limited by the temperature that the magnets reach, which is ideally kept below about 60$^{\circ}$C. With a higher temperature tolerance, the SmCo magnets will allow for longer discharges. In turn, good vacuum pumping in conjunction with longer discharges will remove more neutrals, leading to higher ionization fractions. Additionally, the four-fold increase in magnetic field strength will reduce the cusp loss width by a factor of four. The cusp width can be estimated as $w=4\sqrt{\rho_{i}\rho_{e}}\propto B^{-1}$, where $\rho_{i}$ and $\rho_{e}$ are the ion and electron gyroradii, respectively \citep{Cusp}.

\begin{figure}
  \centerline{\includegraphics[scale=1]{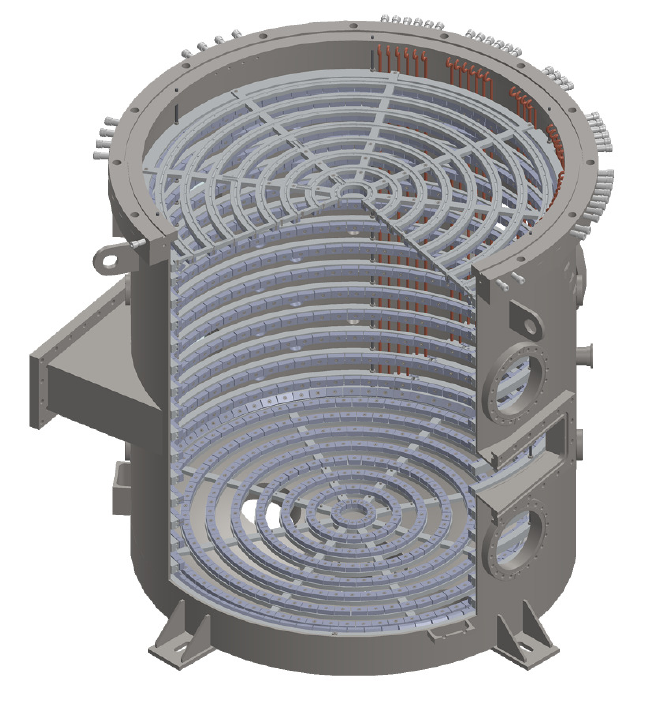}\includegraphics[scale=1]{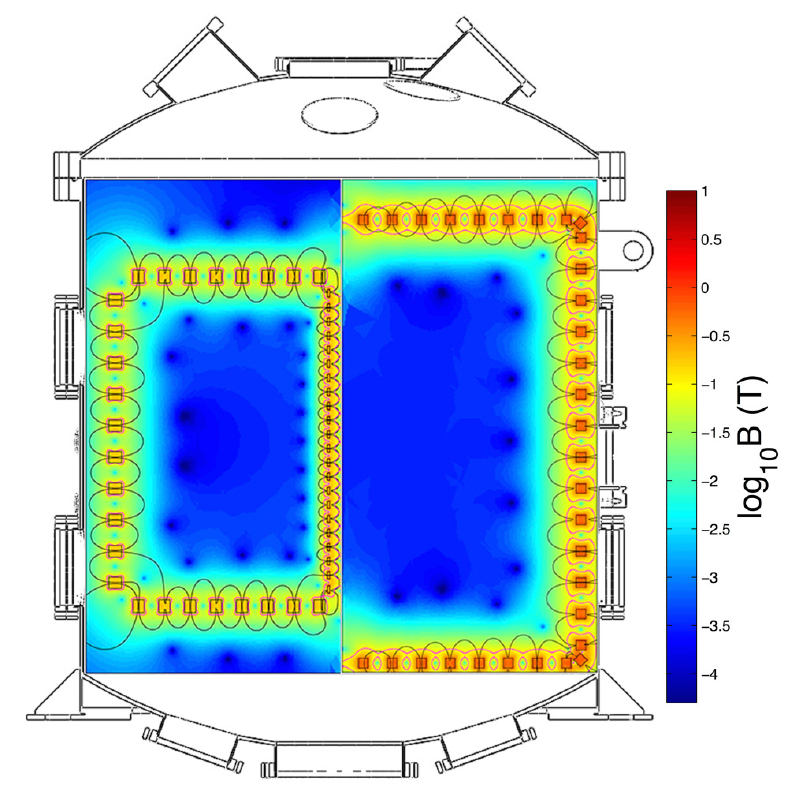}}
  \caption{Left: A design image of the proposed new magnet assembly on PCX-U. Right: A comparison of the current PCX magnet geometry (left) and the proposed geometry (right). These plots show the magnitude of the magnetic field on a scale from Earth's field to 1 T. The pink contour is the 875 Gauss ECH resonance.}
\label{fig:Figure4}
\end{figure}

The left image of Fig. \ref{fig:Figure4} shows a design image of the new magnet assembly. Approximately 1000 magnets are bolted to 14 individually water cooled aluminum rings equally spaced in the axial direction. At the top and bottom of the chamber, 8 concentric rings attached by spokes carrying water cooling support an additional 350 magnets each. Finally, two rings pitched at 45$^{\circ}$ are supported in the corners between the top and bottom rings and the side rings. These angled rings serve to strengthen the field at the corners where the current assembly suffers from the most magnet heating. The whole assembly is supported by 6 rods attached to a top flange which doubles as the water cooling vacuum feedthrough. The inner three rings on the top and bottom assemblies may be removed in order to place future centerstack assemblies of up to 16in in diameter into PCX-U. 

When describing confinement a good figure of merit is the ratio between plasma volume and loss area, representing the particle balance between volumetric ionization and particle flux to loss areas. For this figure of merit, PCX-U represents a significant improvement over PCX. On the right in Fig. \ref{fig:Figure4}, a side by side comparison of the current and new magnet assemblies highlights the roughly 150\% increase in plasma volume (i.e. PCX-U is 1.5 times the volume of PCX). Due to the extra rings of magnets, the cusp length of the new assembly is roughly 130\% longer, but the stronger magnets make the cusp width smaller by a factor of four, so the $\mathbb{V}/A_{\ell}$ ratio is about 4.5 times larger on PCX-U. This favorable improvement will increase ionization fractions in PCX-U significantly. 

In addition to the new magnet assembly, PCX-U includes a second 6 kW magnetron for increased plasma heating, doubling the available microwave power to 12kW. At densities above, $n=7.4\times10^{10}$ cm$^{-3}$, PCX plasmas are above the O-mode cutoff for 2.45 GHz, yet it has been observed that microwave power still heats the plasma possibly via surface wave discharges \citep{CamiPoP}. PCX-U is estimated to operate in this overdense mode. Additionally, the stronger field on the magnets will place the 875 Gauss ECH resonance further into the plasma volume, thus the surface wave discharge will have a larger volume to fill. This extra power will be managed by the high temperature tolerance of the SmCo magnets. 

\section{Model for Global Stability Analysis}\label{sec:Stab}
Here we present a model for studying the global stability of flows in PCX-U as a method for determining the threshold for MRI. Previous global stability analyses for PCX have included the Hall term, but not the neutral drag body force \citep{Ebrahimi2011}. In order to capture both of these effects, an incompressible dissipative Hall MHD model with the neutral body drag force is used. The only two-fluid effect included is the Hall term. Ambipolar diffusion is negligible in PCX since recently charge-exchanged neutrals have a mean free path much longer than the machine size and thus are not coupled to the ions. The neutral drag body force is included as a momentum sink term dependent on the momentum diffusion length. While the incompressibility assumption may not be completely valid near the inner cylinder, we leave detailed study of this effect for future work and assume for this analysis a uniform density. The governing equations used for this analysis are:
\begin{eqnarray}
\frac{\partial {\bf V}}{\partial t}=-({\bf V}\cdot\nabla){\bf V}-\nabla \frac{P}{\rho} +\frac{1}{\mu_{0} \rho}(\nabla \times {\bf B})\times{\bf B}+\nu\nabla^{2}{\bf V}-\frac{1}{\tau_{i0}}{\bf V}\label{eq:MHD}\\
\nabla\cdot{\bf V}=0\\
\frac{\partial {\bf B}}{\partial t}=\nabla\times\left[{\bf V}\times{\bf B}-\frac{1}{\mu_{0} n e}(\nabla\times{\bf B})\times{\bf B}\right]+\eta\nabla^{2}{\bf B}\\
\nabla\cdot{\bf B}=0\label{eq:MHD_end}
\end{eqnarray}
\\
where $P$ is the scalar pressure, $\rho$ is the mass density and $\eta$ is the magnetic diffusivity (in m$^{2}$ s$^{-1}$). In these equations, plasma parameters $n_{e}$, $\rho$, $\tau_{i0}$, $\nu$ and $\eta$ are assumed to be constant and uniform throughout the volume. Profiles of $T_{e}$ and $n_{e}$ from Langmuir probes and the OES system support this assumption \citep{CamiPoP}. The equilibrium modified TCF profile (\ref{eq:TCF}) results from the steady-state balance between the last two terms in (\ref{eq:MHD}). In order to study linear stability these equations are cast in terms of non-dimensional parameters and linearized near the equilibrium state: ${\bf V}_{eq}=V_{\phi}(r)\,{\bf e}_{\phi}$ and ${\bf B}_{eq}=B_{0}\,{\bf e}_{z}$. The unit of length is the radius of the inner cylindrical boundary $R_{1}$ and the unit of velocity is the plasma velocity at this boundary, so ${\bf V}=V_{1}{\bf v}$ and $P=\rho V_{1}^{2}p$. The magnetic field is normalized by the applied axial field: ${\bf B}=B_{0}{\bf b}$. The dimensionless equations are
\begin{eqnarray}
\label{lin_v}
\gamma {\bf v}=-\left({\bf v}_{eq}\cdot \nabla\right){\bf v}-({\bf v}\cdot\nabla){\bf v}_{eq}-\nabla p+\frac{1}{M_{A}^{2}}(\nabla\times{\bf b})\times{\bf b}_{eq}+\frac{1}{Re}\left(\nabla^{2}-\mu^{2}\right){\bf v}\\
\nabla\cdot{\bf v}=0\\
\label{lin_b}
\gamma {\bf b}=\nabla\times\left[{\bf v}_{eq}\times{\bf b}+{\bf v}\times{\bf b}_{eq}-\frac{\delta_{i}}{M_{A}}(\nabla\times{\bf b})\times{\bf b}_{eq}\right]+\frac{1}{Rm}\nabla^{2}{\bf b}\\
\label{end_lin}
\nabla\cdot{\bf b}=0
\end{eqnarray}
where $\gamma$ is the growth rate in units of angular frequency $\Omega_{1}\equiv V_{1}/R_{1}$. The dimensionless parameters that enter these equations are: the fluid Reynolds number, $Re\equiv V_{1}R_{1}/\nu$; the magnetic Reynolds number, $Rm\equiv V_{1}R_{1}/\eta$; the Alfv\'{e}n Mach number, $M_{A}\equiv V_{1}/V_{A} \equiv V_{1}\sqrt{\mu_{0}\rho}/B_{0}$; the normalized ion inertial length for the Hall effect, $\delta_{i}\equiv d_{i}/R_{1}\equiv c/(\omega_{pi}R_{1})$; and the normalized momentum diffusion length for the neutral collision effect, $\mu\equiv R_{1}/L_{\nu}\equiv R_{1}/\sqrt{\tau_{i0}\nu}$. 

Note that in this stability analysis the effect of the neutral drag enters consistently both through the modification of the equilibrium rotation profile and as a drag force in the linearized momentum equation. Equations (\ref{lin_v})-(\ref{end_lin}) are solved for axisymmetric modes using a standard finite difference method assuming no-slip, non-conducting side walls and periodicity in the axial direction. 
We do not take into account the multicusp magnetic field, details of the plasma confinement and driving near the walls. For TCF study we assume that the boundary toroidal velocities are given. By assuming axial periodicity we also ignore the presence of top and bottom endcaps, thus neglecting the possibility of the Ekman circulation and Hartmann layers. These assumptions greatly simplify present analysis, allowing us to focus on the global MRI physics and not on the boundary effects.

The neglected boundary effects are clearly of great importance in the MRI experiments (e.~g., \cite{Gissinger2012}), but they deserve a separate detailed study. In particular, correct boundary conditions near the multicusp edge must be determined by values of viscosity and resistivity in that magnetized region. Empirical measurements and PIC simulations will be needed in order to understand this rather complicated non-MHD boundary condition. In our current model viscosity and resistivity are spatially uniform and calculated for unmagnetized plasma core. Nonetheless, we expect that such approximation still gives us reliable estimates for the parameters required for observing the MRI in PCX. 

\begin{figure}
  \centerline{\includegraphics[scale=1]{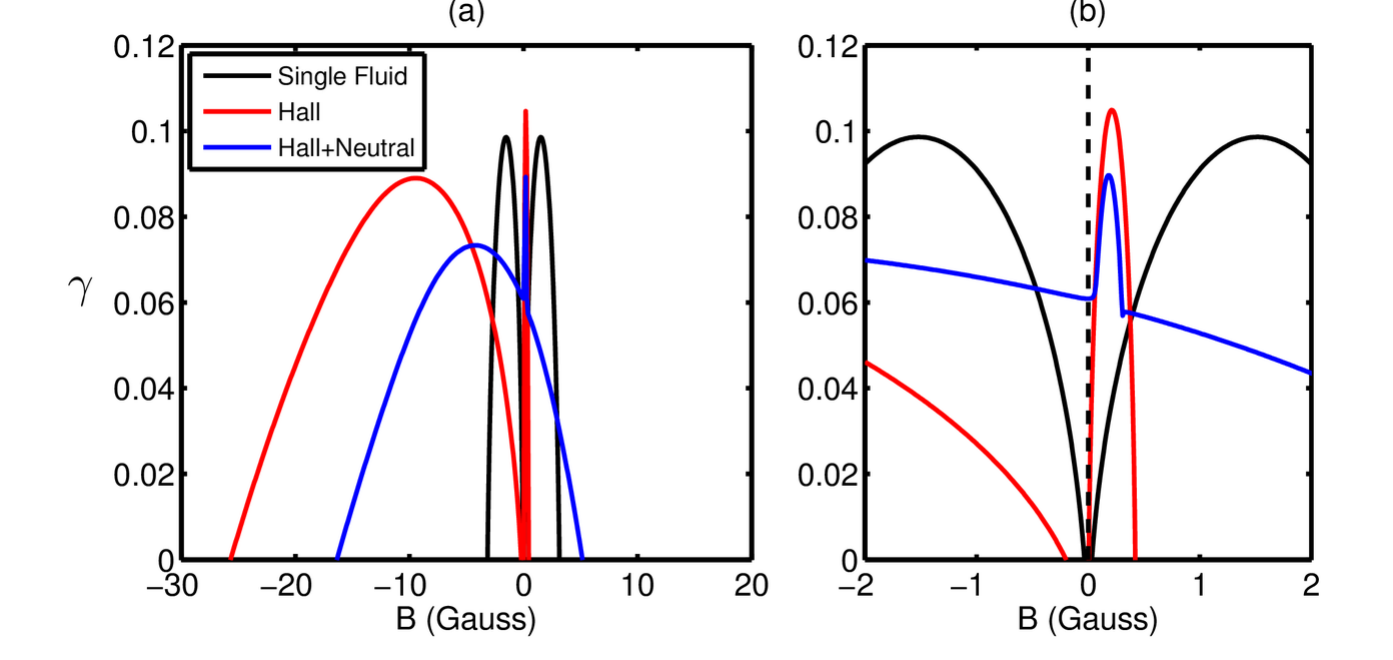}}
  \caption{Growth rate, $\gamma$ in units of $\Omega_{1}$, plotted as a function of applied magnetic field. The different curves present the single fluid MHD case, the inclusion of the Hall term and the inclusion of neutral charge exchange collisions. For this plot $n_{e}=10^{12} cm^{-3}$ and $P_{0}=10^{-5}$ torr corresponding to $\delta=4.55$ and $\mu=0.95$. (a) The full range of magnetic field that gives positive MRI growth rates. (b) A close-up view near $B_0=0$ shows the small (less than earth field) positive $B_{0}$ branch for the Hall and Hall+Neutral cases.}
\label{fig:Figure5}
\end{figure}

\section{Magnetorotational Instability in Taylor-Couette Flow}\label{sec:Stab_results}

The results of the global stability code from the previous section are shown below. Unless otherwise noted, all the analysis in this section is done assuming a singly ionized ($Z=1$ and $n_{e}=n_{i}\equiv n$) helium plasma with $T_{e}=12$ eV and $T_{i}=0.4$ eV. Fixing the temperatures allows viscosity to vary only with density and resistivity, $\eta$, to be mostly fixed (there is a weak density dependence). The dimensions of this system are $R_{1}=0.1$ m, $R_{2}=0.4$ m, and $H=0.8$ m, where the height determines the axial wave-number $k_{z}=2\pi/H$. A neutral-modified TCF profile, as defined in (\ref{eq:TCF}), is used as an initial equilibrium profile for this analysis. The boundary flow velocities are chosen to give a $v_{\phi}\propto1/r$ profile when no neutrals are present. A $v_{\phi}\propto1/r$ profile marginally meets the Rayleigh criterion (but is fully stable when viscosity is included) and meets the ideal-MRI condition. For this analysis an inner velocity of $V_{1}=10$ km s$^{-1}$ was chosen, which sets $V_{2}=2.5$ km s$^{-1}$ when a $v_{\phi}\propto1/r$ profile is desired. All of these fixed values fall into the range of expected parameters and flows for PCX-U.

\begin{figure}
  \centerline{\includegraphics[scale=1]{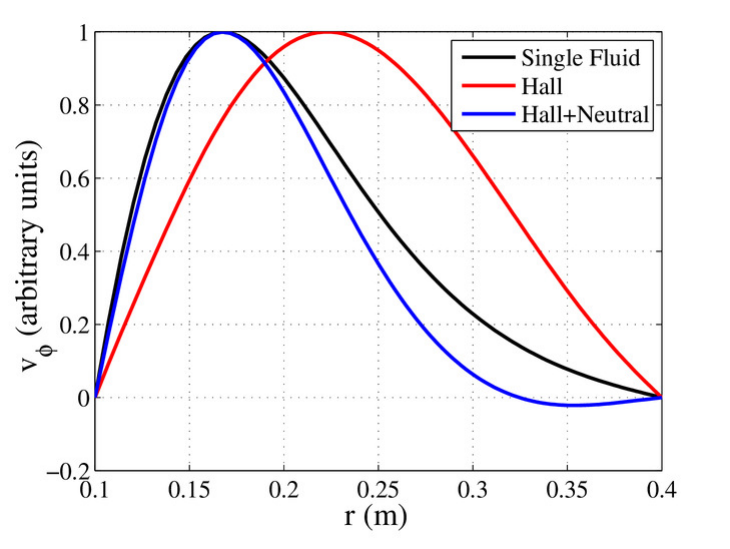}}
  \caption{Toroidal velocity profiles of the most unstable eigen-modes in the Taylor-Couette flow. Corresponding normalized growth rates are $\gamma=0.087$ for single fluid MHD, $\gamma=0.043$ for Hall MHD, $\gamma=0.063$ for Hall MHD with neutrals.  Plasma  parameters are the same as in Fig. \ref{fig:Figure5}. The magnetic field is $B_0=-2$ Gauss and the mode numbers are $m=0$ and $k_{z}=2\pi/H$.}
\label{fig:Figure6}
\end{figure}

The Hall and neutral drag momentum sink terms both produce large qualitative effects on the stability of flows as shown in Fig. \ref{fig:Figure5} (corresponding eigen-modes are shown in Fig. \ref{fig:Figure6}). For the case when neither of these terms are included (single fluid MHD) positive MRI growth rates occur for very small magnitude seed fields (on the order of Earth's field), but the field orientation with respect to the axis of rotation does not matter. When the Hall term is included, positive MRI growth rates are found for stronger seed fields and only when the field is antiparallel to the axis of rotation (negative values of $B_{0}$ in this analysis). If the neutral drag term is added as well, the growth rate is slightly reduced and the magnetic field at the peak growth rate is smaller in magnitude. In the case of all terms being included, increased shear in the modified velocity profile drives a hydrodynamic instability (positive growth rate at $B_{0}=0$) at the particular plasma density and pressure shown in Fig. \ref{fig:Figure5}.

When the dimensionless parameter, $\mu\equiv R_{1}/L_{\nu}$, gets large the neutral-drag modified TCF profile becomes hydrodynamically unstable. Under this condition, the momentum injected at the inner and outer boundaries does not couple across the whole profile. This leads to increased shear at the boundaries of the flow. If the shear is great enough, the Rayleigh criterion is violated for a portion of the profile near the inner boundary, causing the hydrodynamic instability (instability at $B_{0}=0$). This analysis attempts to find a region in parameter space where this hydrodynamic instability is not present ($\mu$ is sufficiently small) and the MRI can still be excited with a weak $B_{0}$. 

\begin{figure}
  \centerline{\includegraphics[scale=1]{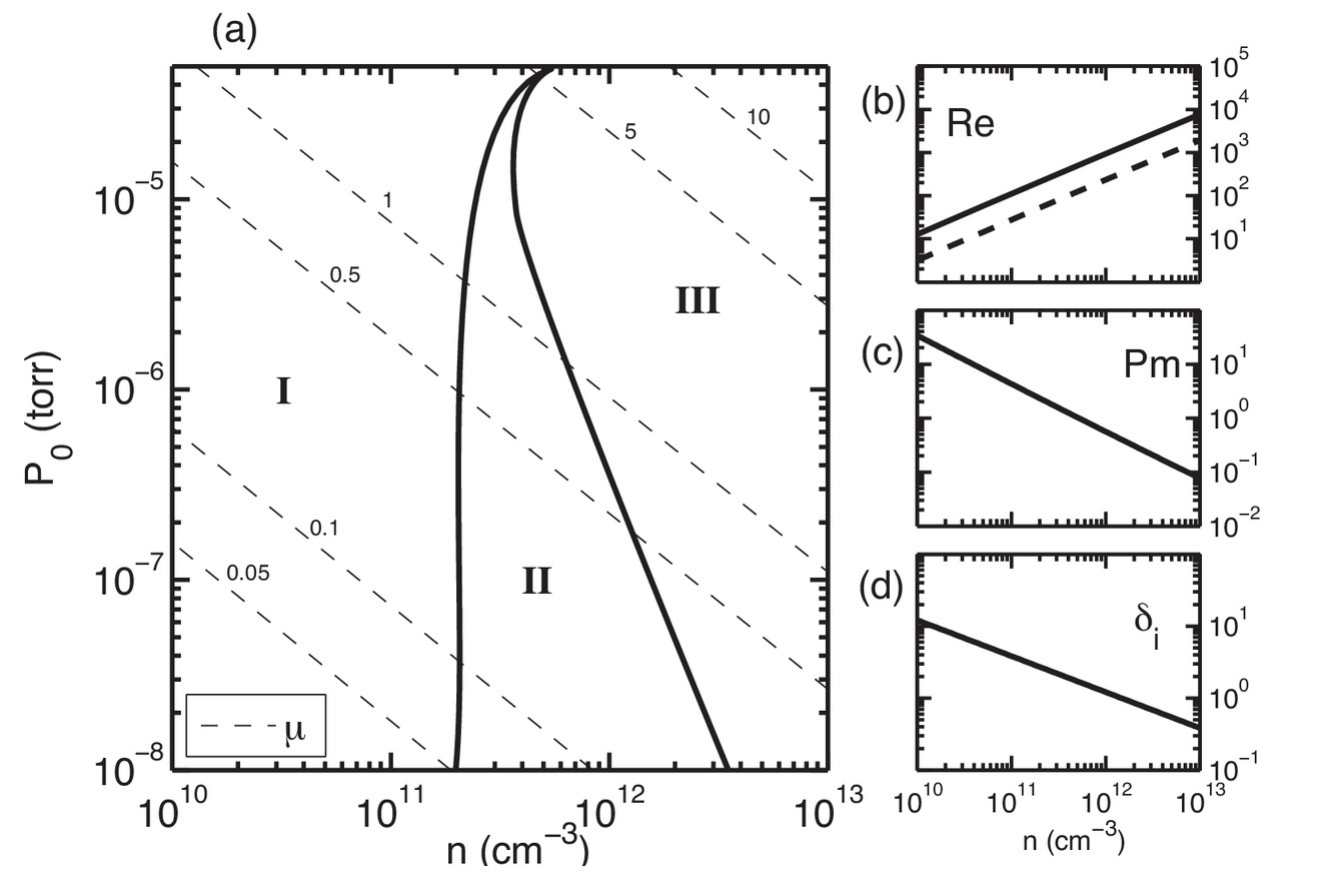}}
  \caption{(a) Stability curves as functions of $n$ and $P_{0}$. For each $n$ and $P_{0}$ the maximum growth rate in the range $B=-300$ to $100$ Gauss is used to determine stability. Region I is stable. Region II is hydrodynamically stable, but unstable to the MRI. Region III is hydrodynamically and MRI unstable. Contours of $\mu$ are also plotted. On the right, the density dependence of (b) $Re$ for both $V_{1}$ (solid) and $V_{2}$ (dashed), (c) $Pm\equiv Rm/Re$, and (d) $\delta_{i}$ are plotted to highlight how relevant dimensionless parameters scale with density.}
\label{fig:Figure7}
\end{figure}

\begin{figure}
  \centerline{\includegraphics[scale=1]{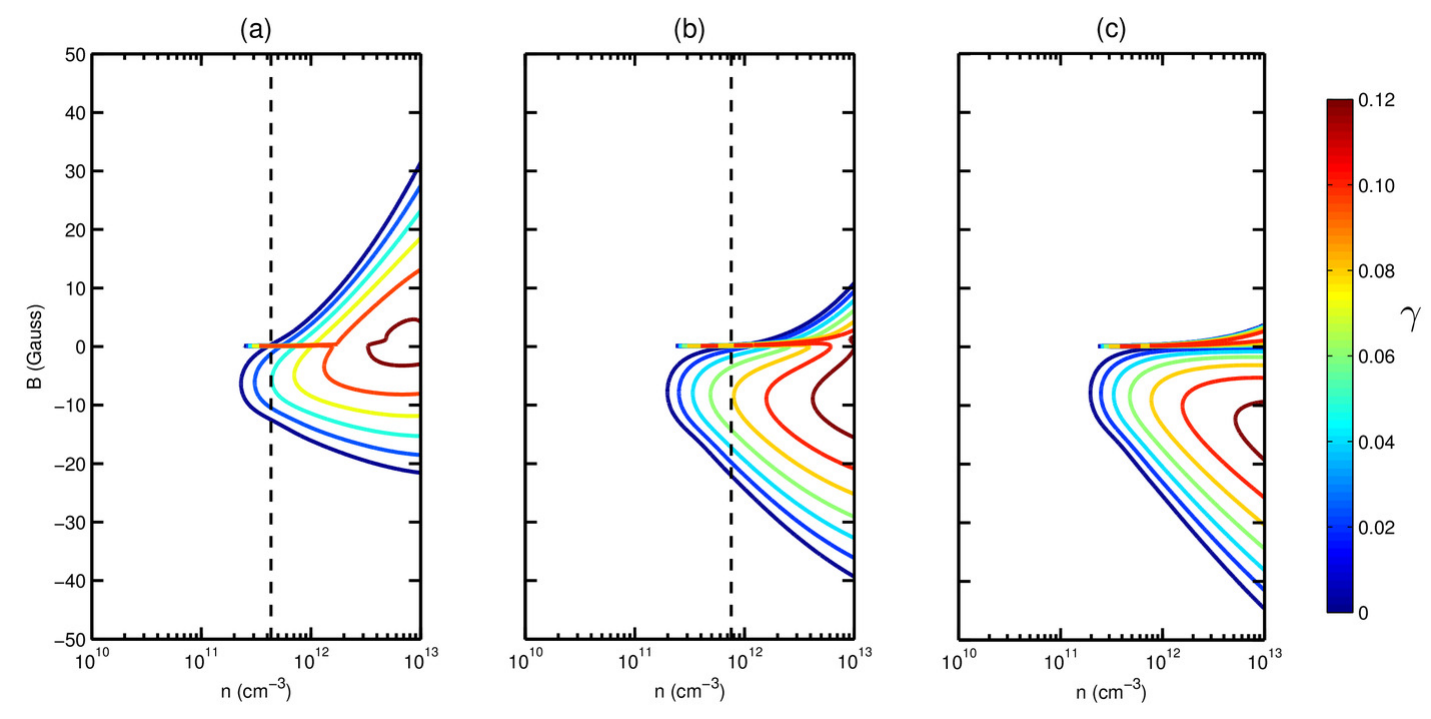}}
  \caption{Contours of normalized growth rate $\gamma$ as a function of density and applied magnetic field for the case with (a) $P_{0}=10^{-5}$ torr, (b) $P_{0}=10^{-6}$ torr and (c) no neutrals. The dashed vertical line represents the density at these neutral pressures above which the flow becomes hydrodynamically unstable due to a decrease in viscosity and increased shear caused by neutral drag.}
\label{fig:Figure8}
\end{figure}

\subsection{Parametric Study of Stability}

By fixing, $T_{e}$, $T_{i}$, $V_{1}$, $V_{2}$ and the dimensions of the system, the remaining variable plasma parameters are the plasma density, $n$; the neutral pressure, $P_{0}$; and the magnetic field $B_{0}$. Such a scan provides a range over which plasma physics phenomena can affect the onset of the MRI. Given these three variables, the dimensionless parameters of interest to this analysis have the following dependencies:
\begin{eqnarray}
&&\delta_{i}\propto n^{-1/2}\\
&&\mu\propto (nP_{0}\ln{\Lambda_{ii}})^{1/2}\\
&&M_{A}\propto n^{1/2}B_{0}^{-1}\\
&&Re\propto n\log{\Lambda_{ii}}\\
&&Pm\propto (n\log{\Lambda_{ii}}\log{\Lambda_{ei}})^{-1}
\end{eqnarray}
where $\log{\Lambda_{ii}}$ and $\log{\Lambda_{ei}}$ are the weakly density dependent Coulomb logarithms for ion-ion and electron-ion collisions, respectively. 

In neutral pressure-density phase space, regions of stability can be mapped out with respect to the MRI and hydrodynamic instabilities as shown in Fig. \ref{fig:Figure7}. Region II represents the region where a MRI experiment would need to operate. Here flowing plasmas are hydrodynamically stable, but an applied axial magnetic field sets off the MRI. It is clear that to ensure that an experiment is in region II with these fixed temperatures and flow velocities, the neutral pressure must be as low as possible and the density must be neither too low nor too high. At higher neutral pressures and higher densities, $\mu$ can become of order unity, at which point the shear in the velocity profile caused by neutral drag is large enough to trigger hydrodynamic instabilities (region III). The MRI threshold between region I and region II appears to be set mainly by the density, i.e. viscosity. For low enough densities, the viscosity is large enough to damp out any instabilities. PCX currently operates in region I, but by boosting confinement (increasing the volume to loss-area ratio), PCX-U is expected to fall inside region II. 

The onset of the hydrodynamic instability can be seen when scanning density at a fixed neutral pressure as in Fig. \ref{fig:Figure8}. For a given neutral pressure there is a density (i.e. viscosity) at which the shear caused by neutral drag becomes great enough to drive a hydrodynamic instability. As the amount of neutrals are decreased this threshold density becomes larger, because less shear is caused by neutrals. In the case when there are no neutrals present (plot (c) in Fig. \ref{fig:Figure8}), there is no hydrodynamic stability and larger MRI growth rates can be reached by increasing the density (lowering the viscosity). With the improved confinement in PCX-U the effect of neutrals can be reduced leading to conditions where the viscosity can be low enough to allow the MRI while still maintaining hydrodynamic stability. 

\subsection{Nonlinear Saturation and Detectability}

\begin{figure}
  \begin{center}
  {\includegraphics[scale=1]{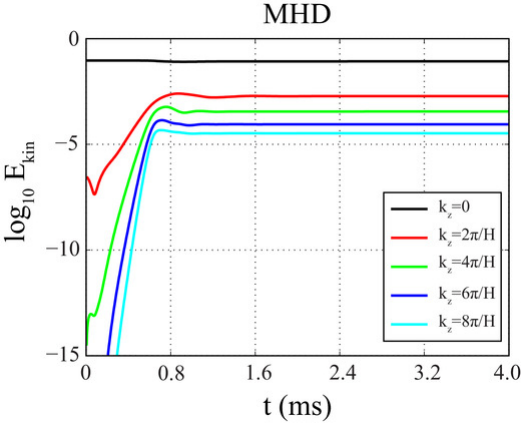}\includegraphics[scale=1]{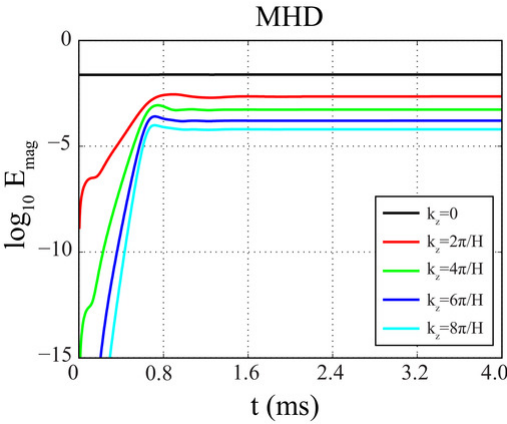}} \\
  {\includegraphics[scale=1]{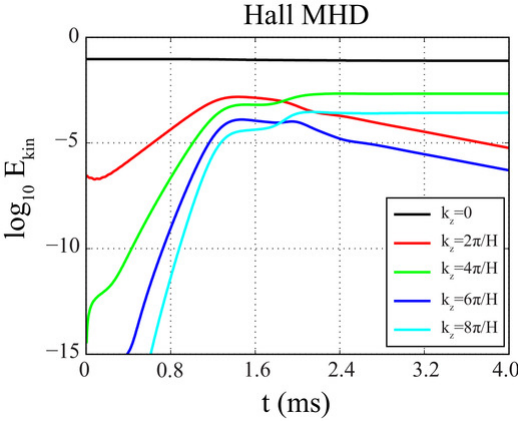}\includegraphics[scale=1]{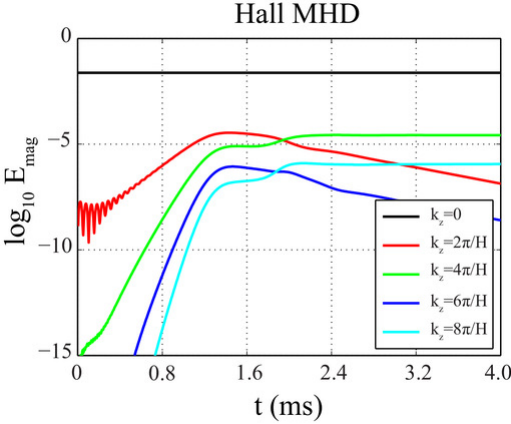}} 
  \end{center}
  \caption{Time evolution of the magnetic and kinetic energies for both the single fluid and Hall-MHD cases.}
\label{fig:Figure9}
\end{figure}

\begin{figure}
\begin{center}
  {\includegraphics[scale=1]{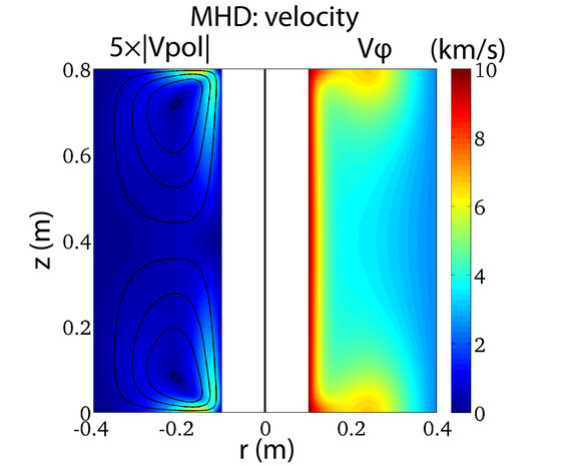}\includegraphics[scale=1]{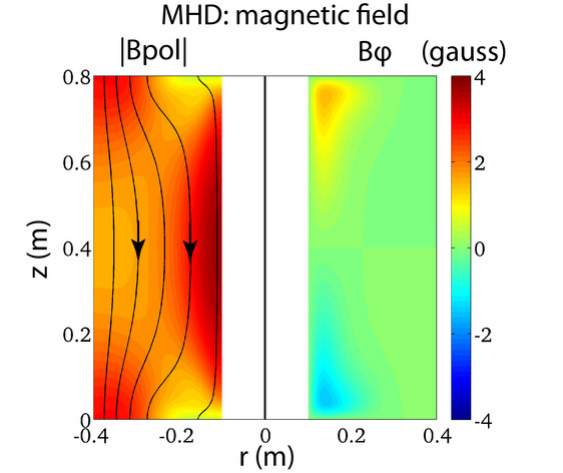}} \\
  {\includegraphics[scale=1]{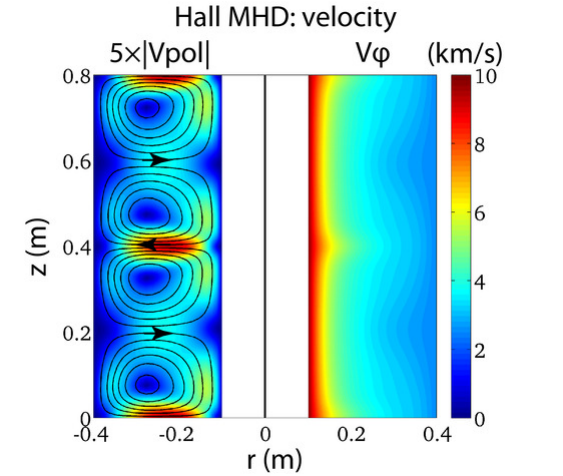}\includegraphics[scale=1]{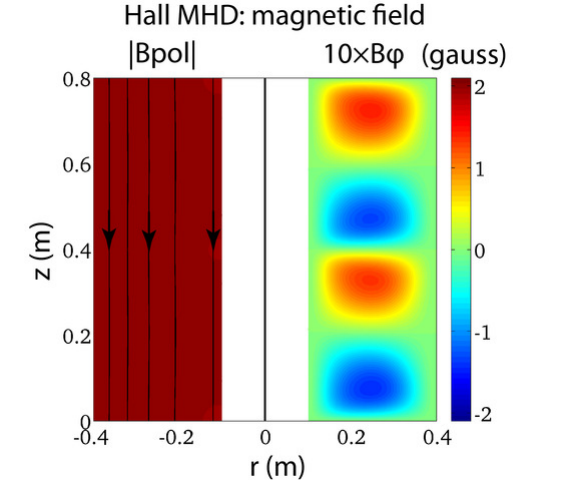}} 
  \end{center}
\caption{Saturated structure of the velocity and magnetic fields for $n_{e}=10^{12}$ cm$^{-3}$ with no neutrals in both the single fluid and Hall-MHD cases. The applied field is $B_{0}=-2$ Gauss. }
\label{fig:Figure10}
\end{figure}

In order to show the saturated state and determine the detectability of the MRI a nonlinear analysis following the model presented above was carried out. Fixing the magnetic field at $B=-2$ Gauss, density at $n_{e}=10^{12}$ cm$^{-3}$ and assuming no neutrals, the full nonlinear system (\ref{eq:MHD})-(\ref{eq:MHD_end}) is evolved in time for single-fluid and Hall MHD cases. Figure \ref{fig:Figure9} shows the kinetic and magnetic energy time-dynamics for these two respective cases and Fig. \ref{fig:Figure10} shows the structure of velocity and magnetic field in the saturated phase of MRI for single-fluid MHD and Hall MHD cases. 

The interesting point here is that the saturation level of magnetic energy for non-zero $k_{z}$ is much lower in the Hall MHD case than in the MHD case. This difference in the energy levels can be explained by noticing that in the Hall MHD case for $\delta_{i}\gg1$ the saturation of the magnetic field occurs when the Hall term is balanced by the induction term in Eq.~(3.3), so $\tilde{b}\sim\tilde{v}/\delta_{i}$, where tilde denotes the parts of magnetic and velocity fields with non-zero $k_{z}$. In the MHD case, the saturation is due to a balance in Eq.~(3.1), so $\tilde{b}\sim\tilde{v}$ and the fields are in equipartition. Also the odd axial harmonics ($k_{z}=2\pi/H,~6\pi/H,\ldots$) turn out to be stable in the saturated Hall MHD state; they are decreasing exponentially in time. 

The saturated magnetic field in Hall MHD is not much different from the originally applied vertical field, as should be expected from consideration of energy evolution plots. However, in the Hall MHD saturated state strong (up to 2 km s$^{-1}$) equatorial jets are produced. Because of lower $Re$ than similar liquid metal MRI experiments, we expect parasitic flows to very minimal, thus allowing this radial jet to be very pronounced. Using either mach probes or the Fabry-Perot system (which is theoretically capable of $\sim10$ m s$^{-1}$ velocity resolution), these strong radial jets will be easily diagnosed, providing evidence of the MRI in PCX-U. Using this signature, onset parameters such as the magnitude of the applied field could be swept to compare to predictions made by this stability analysis. 

\section{Electrically Driven Flow}\label{sec:EDF}
As an alternative to boundary driven TCF, which is greatly affected by the presence of neutrals via the neutral drag body force, we present a scheme for driving flows in PCX-U using a body ${\bf J}\times{\bf B}$ force. This flow scheme, so called electrically driven flow (EDF), is driven by drawing a radial current across the small axial magnetic field provided by external Helmholtz coils. 

EDF in a cylindrical volume produces a $v_{\phi}\propto1/r$ profile in the bulk of the plasma. As opposed to boundary driven flows, the EDF profile shape is unchanged for different driving parameters (total current and ${\bf B}_{0}$). The peak velocity of this profiles is set by the total current that can be drawn radially across the magnetic field. These EDF $1/r$ profiles (as discussed in Sec.~\ref{sec:Stab_results}) are hydrodynamically stable under the Rayleigh criterion when viscosity is included and meet the ideal-MRI conditions, thus constitute a good flow for MRI studies. 

In PCX-U, the Helmholtz coil supplies an axial magnetic field while a single cathode or anode can be placed in the center of the vessel and biased to electrodes on the outer wall to draw the cross field current. The simplicity of this system is experimentally very attractive because an inner boundary electrode assembly is not required as in TCF.  

In order to drive EDF, the current output of PCX cathodes will be increased by replacing the emitting material with lanthanum hexaboride (LaB6). This material has an extremely low work function and when heated is a superb electron emitter, thus making it an ideal candidate for emissive plasma cathodes. LaB6 cathodes used in MPDX routinely draw the maximum current of power supplies ($\sim100$ A), which represents a two order of magnitude increase over the tungsten PCX cathodes \citep{MPDXPOP}. 

\begin{figure}
\centerline{\includegraphics[scale=1]{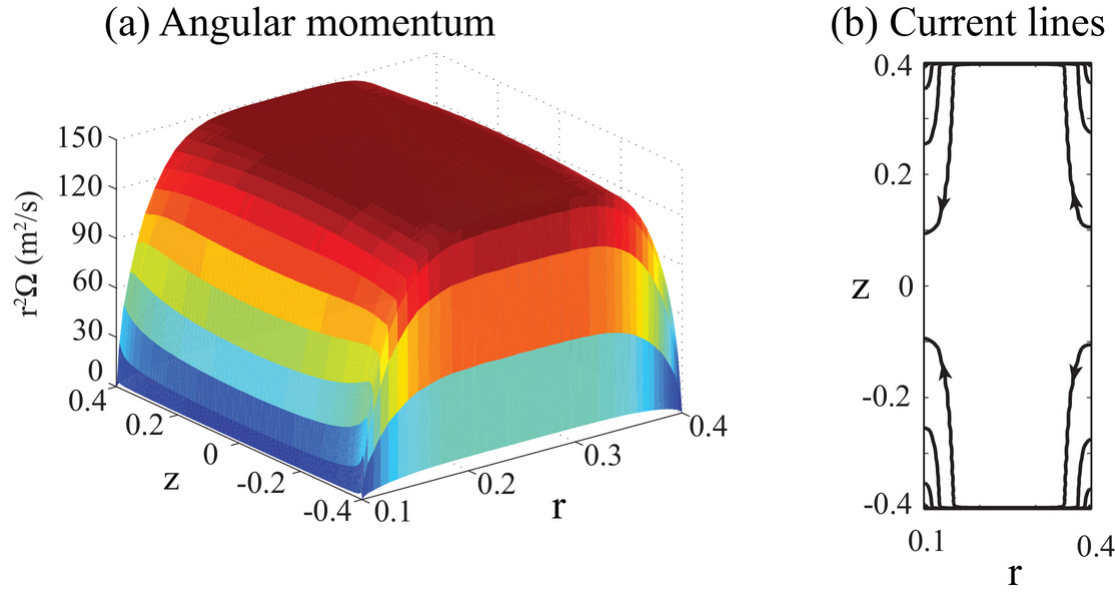}}
\caption{Equilibrium electrically driven flow in single fluid MHD: (a) angular momentum profile, (b) current lines in $r-z$ plane. Narrow Hartmann layers are visible near the top and bottom endcaps.  Calculations are done for a helium plasma with density $n=10^{12}$ cm$^{-3}$, electron temperature $T_e=12$  eV, ion temperature $T_i=0.4$ eV, total radial current $I_0=100$ A and axial magnetic field $B_0=10$ Gauss (corresponding to $Re=18$, $Rm=10$, $\emph{Ha}=385$).}
\label{fig:Figure11}
\end{figure}

\subsection{Equilibrium}
Analysis of EDF in relation to the MRI has been presented in various parameter regimes using dissipative single-fluid MHD \citep{Noguchi_2003,  Khalzov_2006, IvanDeanStab}. In this analysis Hall and neutral collision terms are included to reflect the regime of weakly-ionized sparse plasmas. As an initial step, the axisymmetric equilibrium state ($\partial/\partial t\rightarrow0$) described by the system of equations (\ref{eq:MHD}-\ref{eq:MHD_end}) is found assuming perfectly conducting inner and outer cylinders, insulating top and bottom ends, and no-slip boundary conditions for velocity at the walls. The system is discretized in the two-dimensional $r$-$z$ domain and solved by an iterative method outlined in \citep{IvanDeanStab}.   

\begin{figure}
\centerline{\includegraphics[scale=1]{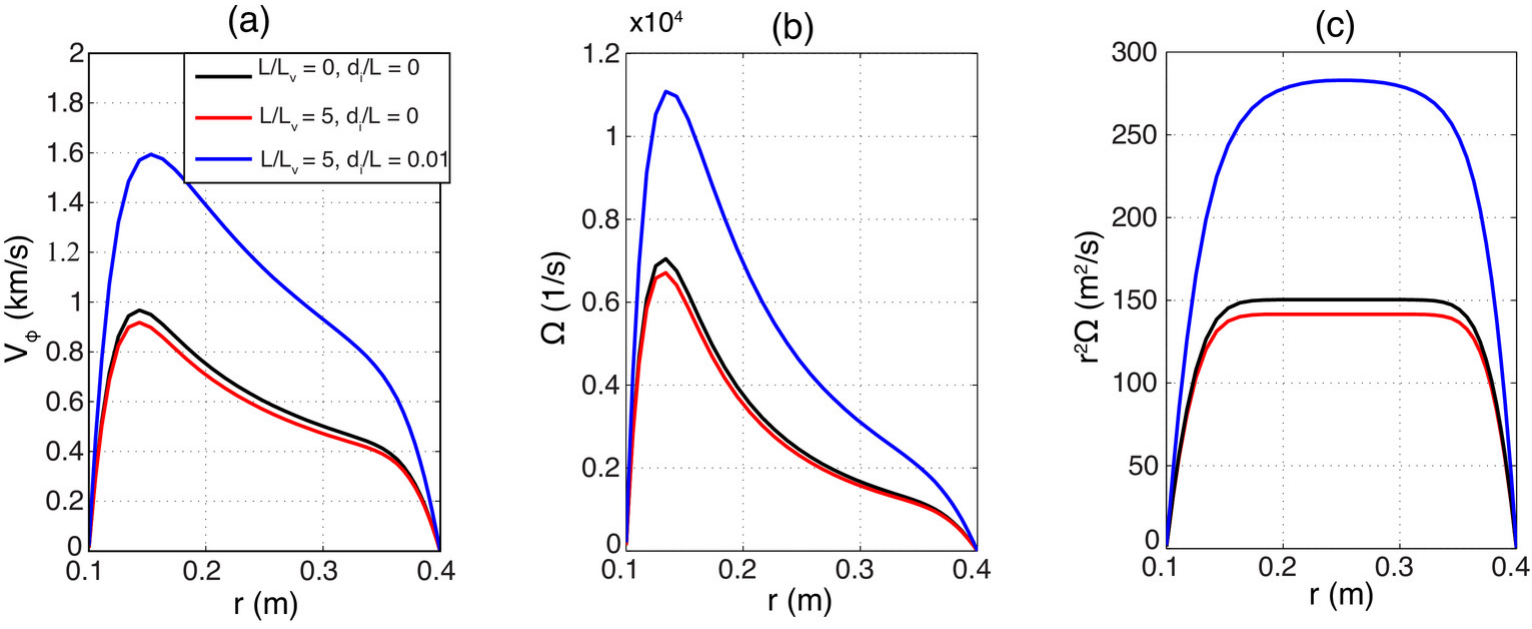}}
\caption{The radial profiles of (a) azimuthal velocity, (b) angular frequency, and (c) angular momentum of electrically driven flow showing the effects of neutral drag and the Hall term. Parameters are the same as in Fig.~\ref{fig:Figure11}.}
\label{fig:Figure12}
\end{figure}

Several key features are present in the equilibrium EDF profiles as shown in Fig. \ref{fig:Figure11}. First, there are thin Hartmann layers near the top and bottom ends, which scale like $1/\emph{Ha}$ (where $Ha\equiv\sqrt{RmRe}/M_{A}$) and parallel layers near side walls, which scale like $1/\emph{Ha}^{1/2}$ \citep{IvanDeanStab}. The poloidal current and associated toroidal field are localized in these boundary layers, which are narrow if $\emph{Ha}\gg1$ and, thus, their presence can be neglected in the following stability analysis. The rest of the plasma rotates with azimuthal velocity $v_\phi\propto1/r$ and is practically current-free. 

Second, since the driving ${\bf J}\times{\bf B}$ force for EDF is a body force applied across the whole profile, the neutral drag body force acts only to uniformly lower the magnitude of the profile. Increased neutral density no longer results in a qualitative change to the shape of the profile like it does for TCF. Because of the fixed profile shape, the bulk EDF is guaranteed to meet the Rayleigh criterion for any finite viscosity. This effectively removes the hydrodynamic instability limit encountered in stability analysis for TCF. Additionally, the Hall term acts to uniformly increase the magnitude of the profile (at fixed total radial current) and does not affect the radial profile shape. This is illustrated in Fig. \ref{fig:Figure12}. 

Third, as shown in \citep{IvanDeanStab}, the magnitude of the secondary flows (poloidal circulation) scales as $Re/\emph{Ha}^2$. At large Hartmann numbers these flows become small, therefore they can be neglected in the stability analysis along with the small boundary layers. Additionally it is assumed that the plasma rotates in the uniform axial field $B_0$ with azimuthal velocity $V_\phi(r)=b/r$, where amplitude $b$ is determined from the equilibrium solver (it depends on total current $I_0$, neutral drag and Hall effect).     

Note that the neglected boundary layers may lead to some additional parasitic instabilities, especially the layer near the outer wall where the angular momentum is decreasing (and therefore violating the Rayleigh criterion). For large $\emph{Ha}$, these layers will be small and localized to a region of lower viscosity in the magnetized cusp region. It is expected that any instability here would not couple into the bulk flow. For completeness, these boundary layers can be removed experimentally by adjusting the positions of the near-wall electrodes so that in addition to bulk forcing, they produce a local drive in the cusp region (like in the TCF) and equalize the boundary and bulk angular momenta. 

\subsection{Stability}
The stability of the equilibrium flow with $V_\phi(r)=b/r$ in an uniform axial field $B_0$ is studied by solving the eigenvalue problem resulting from the linearized system (3.5-3.6). The following results are obtained for singly ionized helium plasma with density $n=10^{12}$ cm$^{-3}$, electron temperature $T_e=12$  eV, ion temperature $T_i=0.4$ eV, and neutral pressure $P_0=10^{-5}$ torr in a cylindrical volume with dimensions $R_1=0.1$ m, $R_2=0.4$ m, $H=0.8$ m. This gives dimensionless parameters, $\mu=0.95$ and $\delta=4.55$.     

\begin{figure}
\centerline{\includegraphics[scale=1]{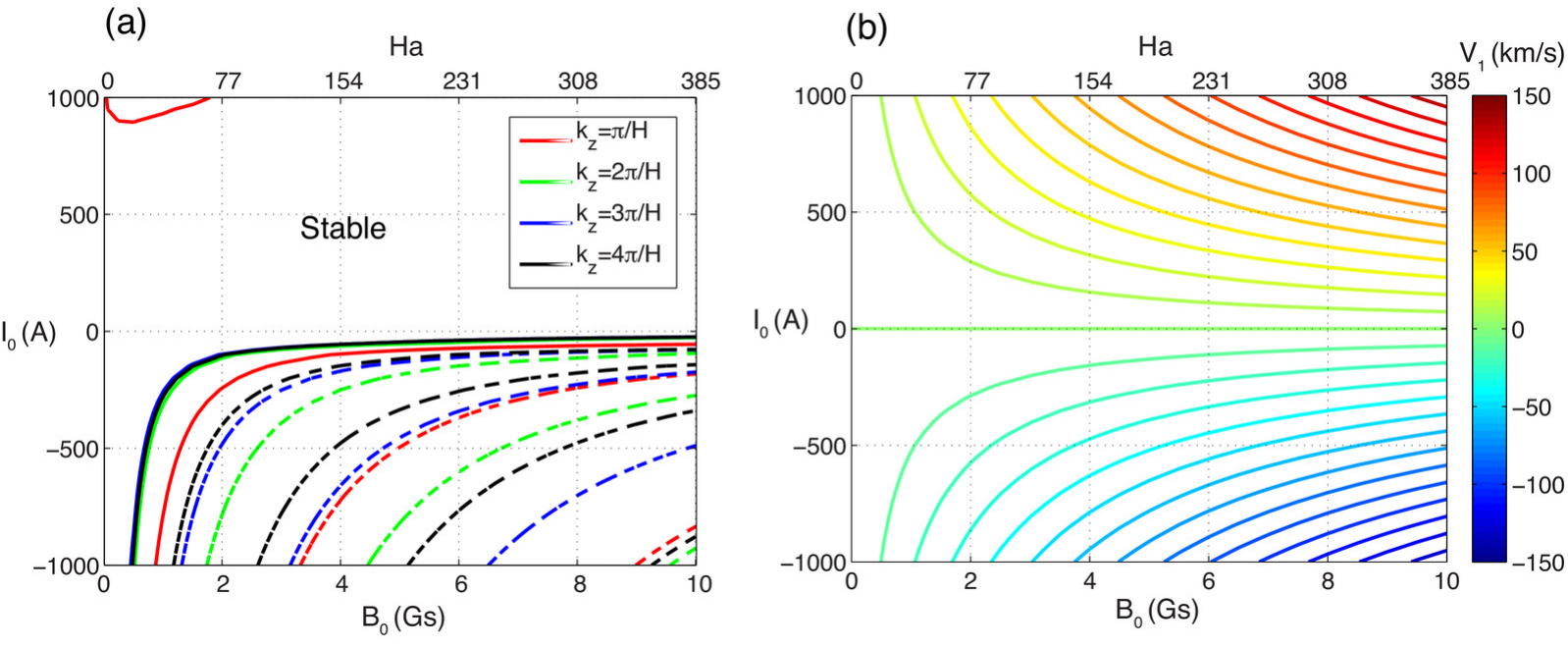}}
\caption{(a) MRI regions in electrically driven flow for modes with different axial wave-numbers $k_z$. Dashed lines of the same color denote stability boundaries for modes with the same $k_z$ but different radial mode numbers.  (b) Azimuthal velocity at the inner wall. Plasma parameters are listed in the text. Negative sign of $I_0$ corresponds to current flowing from the inner to outer wall. }
\label{fig:Figure13}
\end{figure}

The MRI boundary as a function of external field $B_0$ and total current $I_0$ is given in Fig. \ref{fig:Figure13}. The familiar Hall effect is also present in the electrically driven flow where the axis of the flow must be antiparallel to the axial magnetic field to obtain positive MRI growth rates. This means that current must flow from the inner to outer boundary of the plasma. Experimentally, this means that cathodes at the outer edge must be negatively biased with respect to an anode close to ground placed in the center. 

Figure \ref{fig:Figure13} also demonstrates that several modes with different $k_z$ will be excited almost simultaneously when the MRI threshold is reached at total radial currents of $\sim100$ A. Such a multi-mode instability may lead to a fast development of turbulence, which is another intriguing objective for the proposed plasma MRI experiment since in astrophysical systems it is MRI turbulence which leads to enhanced angular momentum transport ultimately. 

\section{Summary and Future Work}
The prospects for observing the MRI in PCX are good. In this paper, the effects of plasma dynamics pertinent to the PCX regime on the MRI have been studied via a global stability analysis. From this study, an experimentally achievable regime has been found where the MRI should be excited in a laboratory plasma. Experimental considerations have led to an upgrade to put PCX in this regime that is currently underway. Finally, so-called electrically driven flow (EDF) is considered as an alternative to boundary driven Taylor-Couette flow and is shown to lead to the MRI with PCX-U parameters. 

A global stability analysis for weakly-ionized sparse plasmas has highlighted the combined effects of ion-neutral collisions and the Hall term on the MRI. As previously noted, the Hall term leads to a requirement that the seed magnetic field be antiparallel to the axis of rotation for positive MRI growth rates. Ion-neutral collisions effectively add a body drag force to the momentum equation. This drag quantitatively changes the profile shape of boundary driven flows and leads to a small region in parameter space where flows are unstable to the MRI while maintaining hydrodynamic stability. 

Finally, an experimental configuration for EDF provides an alternative to boundary driven Taylor-Couette flow that is experimentally attractive due to its simplicity and has a fixed profile shape that is unaffected by the presence of neutrals. EDF can be easily realized in PCX-U with high power LaB6 cathodes capable of driving high $Rm$ flows. Preliminary global stability analysis of EDF profiles shows that the MRI and possibly MRI turbulence can be excited at experimentally achievable parameters. It should be noted that Keplerian flows are body force driven (as is EDF), such that many of the difficulties associated with exciting the MRI in boundary driven flows (such as the neutral drag) are not present in the true astrophysical systems under study. In this way using EDF to study the MRI removes details that are not present in real accretion disks.

We emphasize again that the presented results of the MRI analysis are based on numerical modeling with a number of simplifying assumptions. In particular, we avoid complications related to the presence of top and bottom end-caps by assuming axial periodicity of the system, we simplify the boundary conditions of the magnetized cusp, and we neglect the small boundary layers in EDF and electrical currents flowing in them. We recognize that these are important issues and plan to address them in more detail in future work, especially with respect to determining the proper boundary conditions for the magnetized cusp region. Despite these simplifying assumptions, we believe all the analysis in this paper provides good estimates for MRI onset in PCX that can help to inform the design of PCX-U and other plasma MRI experiments. \\ 

This work was funded in part by the National Science Foundation (NSF) and the Center for Magnetic Self Organization in Laboratory and Astrophysical Plasmas (CMSO). 

\bibliographystyle{jpp}

\bibliography{JPP2014}

\end{document}